\documentclass[preprint,twocolumn,tighten,times]{aastex631}

\usepackage{amsmath}
\usepackage{mathtools}
\usepackage{bm}
\usepackage{xcolor}

\newcommand\gmode{$g$-mode}
\newcommand\msun{\ensuremath{M_\sun}}
\newcommand\teff{\ensuremath{T_{\rm eff}}}

\newcommand{\bvfreq}{Brunt-V\"ais\"al\"a frequency}

\newcommand{\project}[1]{\textsl{#1}}
\newcommand{\gaia}{\project{Gaia}}

\shortauthors{Montgomery \& Dunlap}
\shorttitle{Fluid Mixing in Crystallizing White Dwarfs}

\begin{document}
\title{Fluid Mixing during Phase Separation in Crystallizing White Dwarfs} 

\author[0000-0002-6748-1748]{M. H. Montgomery}
\affiliation{University of Texas and McDonald Observatory, Austin,
  TX, USA}

\author[0000-0002-1086-8685]{Bart H. Dunlap}
\affiliation{University of Texas and McDonald Observatory, Austin, TX, USA}


\begin{abstract}

Accurate models of cooling white dwarfs must treat the energy released as their cores
crystallize. This phase transition slows the cooling by releasing latent heat and also
gravitational energy, which results from phase separation: 
liquid C is released from the solid C/O core, driving an outward carbon flux.
The \gaia\ color-magnitude diagram provides striking
confirmation of this theory by revealing a mass-dependent overdensity of white dwarfs,
indicating slowed cooling at the expected location. However, the observed overdensity is enhanced
relative to the models. Additionally, it is associated with increased magnetism, suggesting 
a link between crystallization and magnetic field generation. Recent 
works aimed at explaining an enhanced
cooling delay and magnetic field generation employ a uniform mixing prescription that assumes large-scale turbulent motions; 
we show here that these calculations are not self-consistent.
We also show that thermohaline mixing is most likely efficient enough to provide the required
chemical redistribution during C/O phase separation, and
that the resulting velocities and mixing lengths are much smaller than previous estimates.
These reduced fluid motions cannot generate measurable magnetic
fields, suggesting any link with crystallization needs to invoke a separate mechanism.
Finally, this mixing alters the chemical profiles which in turn affects the frequencies of the pulsation modes.

\end{abstract}

\section{Astrophysical Context}

The \gaia\ mission has uncovered a clear sequence in the white dwarf
color-magnitude diagram that confirms theoretical expectations \citep{VanHorn68} that
the steady cooling of white dwarfs (WDs) slows as they crystallize
\citep{Tremblay19}. Termed the ``Q-branch,'' this pile-up of high mass 
(0.9 --  1.1$\,M_\odot$) WDs coincides with carbon/oxygen models that have 
begun crystallizing. However, the detailed shape of the observed bump in 
the WD luminosity function is both narrower and higher than predicted by the 
models, indicating an extra cooling delay, possibly up to 8\,Gyr for some 
fraction of the population \citep{Cheng19}. This has led to renewed interest 
in the crystallization of mixtures and in $^{22}$Ne sedimentation and
``distillation'' \citep{Blouin20a,Blouin21a,Bauer20}.

Numerous calculations, using molecular dynamics
\citep{Horowitz10,Schneider12}, density functional theory 
\citep{Segretain93}, and free energy approaches \citep{Blouin20a,Medin10},
support the claim that the different ionic species in the interior of
a white dwarf should partially separate upon crystallization. For
a binary elemental mixture, the high-$Z$ species should be enhanced in
the solid relative to the low-$Z$ species, with the surrounding fluid
being depleted in the high-$Z$ element.
Subsequently, through a Rayleigh-Taylor (RT) instability, the lower-$Z$ (and 
therefore lighter) liquid layer is assumed to mix with the denser layers above
it \citep[e.g.,][]{Isern97,Salaris97}. This chemical redistribution
releases gravitational energy, providing an additional energy source
which slows the cooling of the WD
\citep[e.g.,][]{Stevenson80,Mochkovitch83,Isern97,Salaris97,Montgomery99,Althaus03,Bauer23}. This
mixing also systematically alters the \gmode\ oscillation frequencies
of these stars \citep[e.g.,][]{Montgomery99a,Corsico05a,DeGeronimo19}.

\section{Mixing Re-Examined}

\begin{figure*}[t!]
  \includegraphics[width=0.96\columnwidth]{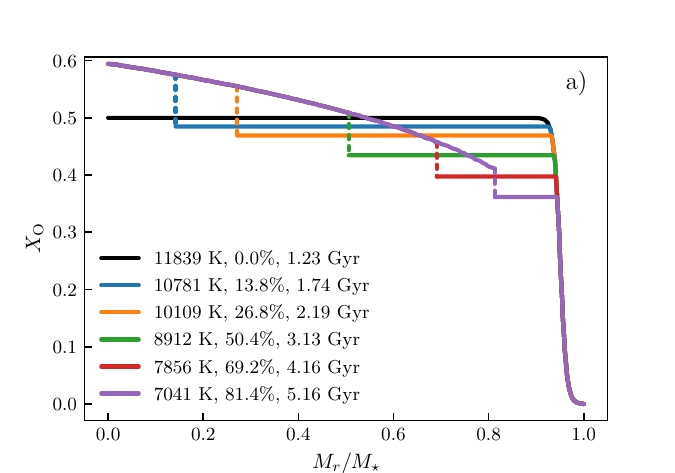}
    \hfill
  \includegraphics[width=1.04\columnwidth]{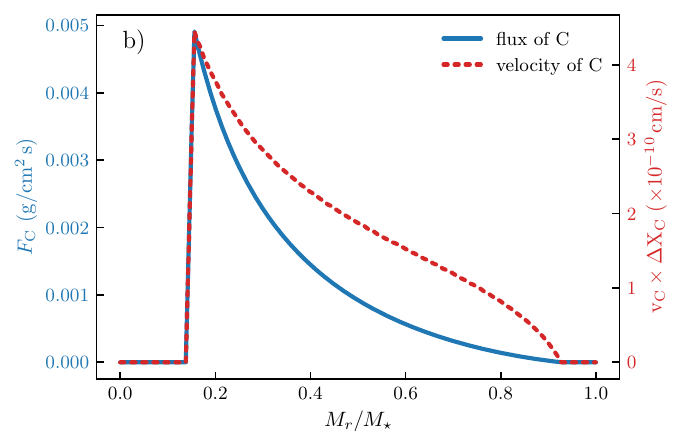}
  \caption{\emph{(a)}
  The evolution of the oxygen profile due to crystallization
    and phase separation for a 1 \msun\ C/O WD. The different curves
    are labeled with the \teff, percent crystallization by mass, and
    age of the model, respectively. Initially, the C/O core has a
    uniform composition out to $M_r \sim 0.9\,M_\star$, with an
    oxygen mass fraction of 0.5. The dotted lines indicate the
    boundary between the solid and liquid regions in each model. The
    carbon profile for $M_r/M_\star < 0.9$ is essentially equal to
    $1-X_{\rm O}$.
    \emph{(b)}: The carbon flux (blue solid curve, left axis) and velocity
    times carbon enhancement
    (red dashed curve, right axis) of carbon-enriched bubbles derived
    from two consecutive evolutionary models at $\sim 16\,$\%
    crystallization by mass.
    \label{xtal_prof}
    \label{F_v}
  }
\end{figure*}

The suggestion has recently been made that the fluid
motions resulting from phase separation could be large enough to generate
magnetic fields of order $\sim\,$0.1$\,$MG \citep{Isern17}, and even larger
fields are estimated in the recent calculations of \citet{Ginzburg22}.
We examine the viability of this hypothesis through estimates of
the mass flux and fluid velocity required in the standard treatment of phase 
separation. We examine different mixing mechanisms and discuss the consequences our 
results  have for magnetic field generation  and pulsations in these stars.

From the perspective of a 1-D stellar evolution code, at the end of a
given time step the model will have increased its crystallized mass
fraction, leaving a thin carbon-enhanced layer on top of the
crystallized core. Since carbon is lighter than oxygen 
at these conditions 
($(\delta \rho/\rho)_0 \sim 3 \times 10^{-3}$), 
this fluid layer will be
RT unstable to mixing. Given the large acceleration
due to buoyancy felt by a pure carbon fluid element
($g \,\delta \rho/\rho \sim 10^{5} \, \rm cm/s^2$), this mixing is
assumed to be rapid and vigorous, producing a layer of uniform
composition out to a point in the model where the carbon mass fraction
matches that of the layer above it. This ensures that the
resulting configuration is stable to further RT mixing. In section~\ref{re-mix}
we explore the case in which the carbon enhancement of a fluid bubble
is much smaller, resulting in correspondingly smaller accelerations.

The detailed mechanics of this mixing are not usually considered, e.g., the
size or velocity of the convective bubbles. Recently, \citet{Isern17}
have examined this mixing in more detail. They assumed bubbles of radius 
$\sim\,$0.1$\,R_{\rm core}$, where
$R_{\rm core}$ is the radius of the crystallized core. Balancing the
buoyancy force with that due to turbulent drag, they computed
velocities of $\sim \,35\, \rm km \,s^{-1}$.  Using relations in
\citet{Christensen10}, they suggested that this could lead to a dynamo
process that could generate magnetic fields of order 0.1~MG.
More recently, \citet{Ginzburg22} have estimated a much smaller carbon
enhancement for the fluid bubbles, resulting in reduced mixing velocities
of order $10^2\,\rm cm \,s^{-1}$.

From evolutionary calculations we know that the rate of crystallization
should be small, with WDs taking $\sim 10^9$ years to crystallize most
of the mass in their cores; at any given point in the crystallization
process, the radius of the crystallized core is increasing at a rate
of approximately $10^{-8}\rm \, cm\, s^{-1}$. Therefore, the thickness
of the carbon-enhanced layer above the crystallized core should be
increasing at a similar rate.  Thus, we expect that the net outward
flux of carbon should be very small. We estimate these numbers in the
following section.

\section{Chemical Transport in the Standard Picture}
\label{chem_trans}

For the calculations described in this and subsequent sections we have 
used version 15140 of the stellar evolution code MESA. In particular,
the numerical details of crystallization and fluid mixing, both neutrally
buoyant and the standard prescription, are described in Appendix~\ref{num}. 
We note that more recent versions of MESA have the ability to do the 
standard uniform mixing during crystallization as a built-in option
\citep{Bauer23}, but we have not used this functionality.

We calculate results of the standard prescription for crystallization
and phase separation in an evolving WD model, assuming that the fluid
mixing leads to a uniformly mixed overlying fluid layer. 
This model has a carbon/oxygen core and a total mass of $1\,M_\odot$. 
We have artificially chosen an oxygen profile that extends past $0.9\,M_\star$
in an effort to maximize the fluid mixing during subsequent evolutionary stages 
and a high mass so that the model will be partially crystallized in the DAV
instability strip. For the phase diagram of carbon/oxygen crystallization,
we use the recent results of \citet{Blouin20a}.
In Figure~\ref{xtal_prof}a we show the evolution of the chemical
profile of oxygen as the model cools. The apparent discontinuity in
the profiles (dashed portion of curves) indicates the edge of the
crystallized core and the fluid layers above it.

\begin{figure*}[t!]
  \centering{\includegraphics[width=0.94\columnwidth]{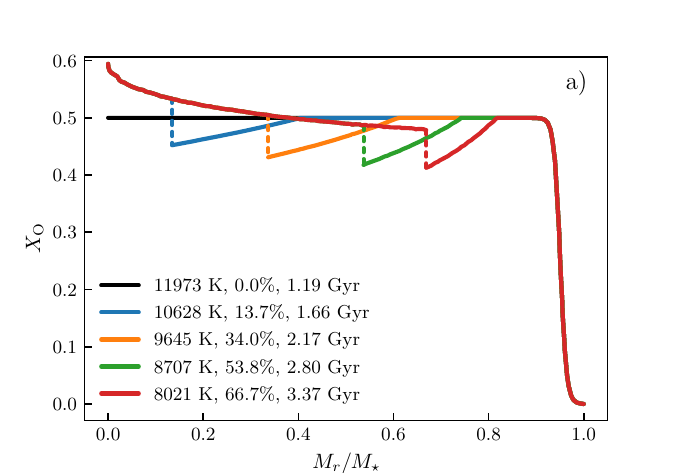}}
  \hfill
  \centering{\includegraphics[width=1.06\columnwidth]{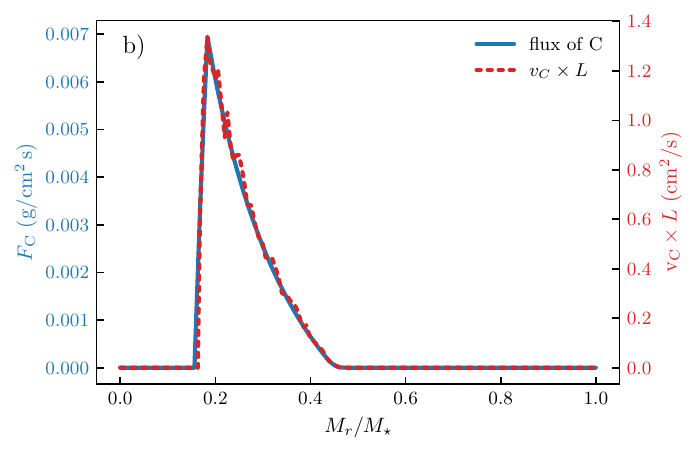}}
  \caption{
  The same as Figure~\ref{xtal_prof}, except for the case of neutrally buoyant
  mixing. \emph{(a)}: Instead of flat profiles in the mixed regions, these
  profiles have a non-zero slope that results in $N^2 \sim 0$. 
  \emph{(b)}: 
  Using equation~\ref{feqn2} we are able to solve for the velocity times the
  mixing length, which is shown on the right-hand axis, again assuming $f=0.5$.
  We note that $v_C \times L \sim 1\rm\, cm^2/\,s$.
    \label{XO_profs}
    \label{xcomp}
    }
\end{figure*}

Using consecutive models with differing crystallized mass fractions,
we can compute the mass flux of carbon as a function of radius that is necessary
to transform the carbon profile at one time step into the carbon profile at the
next time step; we plot this as the solid blue curve (left axis) in
Figure~\ref{F_v}b. We note that when the model is
about 14\% crystallized by mass, the uniformly mixed fluid region
extends from $\sim\,$0.1 to $\sim\,$0.9 in $M_r/M_\star$. This stage
of crystallization was chosen in an effort to maximize the extent of
the mixed region and therefore the fluid velocities.

Next, we compute the magnitude of fluid velocities that are required to
produce this mixing. Assuming that the carbon enhancement of a fluid
bubble relative to its environment is given by $\Delta X_{\rm C}$, and
that the bubbles have
a filling fraction of $f$, then the flux of carbon, $F_C$, is related
to the velocity, $v_{\rm C}$, by
\begin{equation}
  F_{\rm C} = f\,\Delta X_{\rm C}\, \rho \, v_{\rm C},
  \label{feqn}
\end{equation}
where $\rho$ is the mass density. 
Taking $f = 0.5$, we find that the product $v_{\rm C} \times \Delta X_{\rm C}$ 
is given by the red dashed curve (right axis) in Figure~\ref{F_v}b. 
If $\Delta X_{\rm C} \approx 10^{-2}$, then the velocities are quite low, of
order $\sim 10^{-8}\,\rm cm\, s^{-1}$, which is $\sim 10^{14}$
times smaller than the estimate of \citet{Isern17}. On the other hand, if we 
assume values of $\Delta X_{\rm C}$ of $10^{-16}$ and $10^{-12}$, then
we obtain velocities similar to those of \citet{Isern17} and \citet{Ginzburg22}, 
respectively.  Thus, the velocities needed to generate appreciable magnetic fields or 
achieve turbulent mixing require very small values for $\Delta X_{\rm C}$.
However, in the following section, we show that such a small carbon enhancement results 
in bubbles that are limited to small vertical displacements.

\section{Physical Inconsistency}
\label{inconsistency}

The mixing envisioned in \citet{Ginzburg22} and \citet{Isern17} is 
vigorous enough to produce a uniform chemical profile in the mixed
region and is relatively rapid in the sense that the convective turnover time, 
$\tau_c$, is short compared to a bubble's
thermal diffusion time scale, $\tau^{*}_{\rm KH}$ \citep[see][]{Kippenhahn80}. 
To see this, we use 
\begin{equation}
\tau^{*}_{\rm KH} \approx \frac{L^2}{\kappa_T},
\end{equation}
where $L$ is the radius of a spherical bubble and $\kappa_T$ is the thermal diffusivity 
as defined in Appendix~\ref{thermohal}. Through most of its evolution,
the assumed mixing length in the model of Ginzburg et al.\ is given by a pressure scale
height. Taking $L$ as half of this mixing length, we find a thermal diffusion
time scale for the bubble of $\tau^*_{\rm KH}\sim 2 \times 10^{6}\,$yr.
The ratio of
$\tau_c$ 
($\sim 0.1$\,yr)
to $\tau^*_{\rm KH}$ is, then, $\sim 10^{-7}$, meaning that there will be almost no
heat exchange between the bubble and its environment. Thus, the thermodynamics
of the bubble will essentially be adiabatic;
yet, as we will show, this contradicts the assumptions made 
in deriving $\tau_c$.

In uniform composition convection zones in stellar atmospheres, 
the temperature decreases outwards so that when a fluid parcel rises and
expands in response to the lower pressure, the possibility exists for it
to be less dense than  its surroundings and continue to rise.  This will 
happen if, when it cools adiabatically, it is still hotter than its  
surroundings (this is just the  Schwarzschild criterion). 
In contrast, in a  nearly isothermal environment (still with uniform 
composition) such as the core of a WD, a rising element that expands 
adiabatically will be cooler than its surroundings, making it 
denser and hence convectively stable. On the other hand, if the composition 
of the rising parcel is such that it is less dense than the surrounding 
fluid, then it may continue to rise, but the cooling effect on its density 
must still be taken into account. The derivation of convective velocities in 
\citet{Ginzburg22} only considers the effect of composition on density and 
not the effect of adiabatic cooling on density,  which can very quickly 
dominate in a nearly isothermal environment. We quantify this argument below. 

The return force on an adiabatically displaced fluid element in a uniform
composition medium is proportional to $N^2\,\delta r$, where
$\delta r$ is the vertical displacement of the fluid element from its 
equilibrium  position and $N^2$ is the square of the \bvfreq.  
On the other hand, if the fluid element is enhanced in carbon relative to 
its environment, 
it will experience an additional upward buoyancy force proportional to 
$g (\delta \rho/\rho)_0\, \Delta X_{\rm C}$,  where $(\delta \rho/\rho)_0$ 
is the fractional density contrast between a pure oxygen and a pure carbon
plasma ($\sim 3 \times 10^{-3}$ in our models), and $\Delta X_{\rm C}$ is 
the enhancement of the mass fraction of carbon in the fluid element compared
to its surroundings. Including both effects yields
the following differential equation for the acceleration of the
fluid element: 
\begin{equation}
   \frac{\partial^2 \delta r}{\partial t^2} \,\, = \,\, \overbrace{-N^2\, \delta r}^{\rm adiabatic} \quad  + \quad
   \overbrace{g \left( \frac{\delta\rho}{\rho} \right)_0\,\Delta X_{\rm C}}^{\rm composition\,\, driven}.
   \label{acc}
\end{equation}

Since the background chemical profile is assumed to be uniform, 
$\Delta X_{\rm C}$ is independent of the height $\delta r$, so the 
second term on the RHS of equation~\ref{acc} is approximately constant. 
On the other hand, the first term is 
proportional to $\delta r$, so it becomes more negative the farther up the fluid element travels. 

\citet{Ginzburg22} omit the first term on the RHS of equation~\ref{acc} in their analysis, so their fluid elements can rise without bound. This 
naturally leads them to assume a mixing length that is of order a
pressure scale height. However, when we include both terms in equation~\ref{acc}, we find
that a fluid element that starts out at $\delta r = 0$ with a positive
carbon enhancement will eventually rise to a new equilibrium point where 
these two terms balance each other. This equilibrium distance is given by
\begin{equation}
   \delta r_{\rm eq} = \frac{g}{N^2}\left( \frac{\delta\rho}{\rho} \right)_0
    \Delta X_{\rm C}
   \approx L.
    \label{ML_est}
\end{equation}
In other words, the value of $\delta r_{\rm eq}$ in 
equation~\ref{ML_est} is the 
actual mixing length $L$ that should be used based on these assumptions.

Using the values from \citet{Ginzburg22} we estimate their model predicts
$(\delta \rho/\rho)_0\,\Delta X_{\rm C}  \sim 10^{-14}$, and for values 
of $g$ and $N^2$, we use a  fiducial WD model.  We find that $L \sim 
0.03\,$cm. This value is approximately 10 orders of magnitude 
smaller than their assumed mixing length of 
$\sim 10^8\,$cm, highlighting a severe inconsistency. 

To summarize this chain of reasoning, the assumptions of \citet{Ginzburg22} 
of (1) a large mixing length (of order a pressure scale 
height), (2) a bubble size that is of order this mixing length, and (3) 
instantaneous thermal equilibration of the 
rising fluid element with its surroundings (i.e., neglect of 
the adiabatic term in equation~\ref{acc}), lead to fluid 
velocities that imply a convective turnover time that is much 
shorter than the thermal diffusion time scale. This requires 
the resulting fluid motions to be adiabatic, which is 
contrary to (3). Thus, the model of \citet{Ginzburg22} is 
invalid in this context. 

\section{Neutrally Buoyant Mixing}
\label{re-mix}

In this section, we derive conditions whereby weak RT mixing could produce a profile that is very close to neutral stability. By assuming that the fluid motions are adiabatic, we are implicitly ignoring the physics related to thermohaline mixing, which  we examine in detail in sections~\ref{thermo} and \ref{thermo2}. 
The main motivation for this section is to provide a calculation of an upper limit for the steepness of 
the chemical profile. 

The \bvfreq\ can be written in a form that explicitly displays its dependence on the composition profile:
\begin{eqnarray}
\label{bv0}
  N^2 & = & \frac{g^2 \rho}{P} \frac{\chi_T}{\chi_\rho} \left(\nabla_{\rm ad} -
    \nabla + B\right) \\
      & \equiv & N_T^2 + N_\mu^2,
\label{bv1}
\end{eqnarray}
where we have written $N_T$ and $N_\mu$ for the thermal and compositional parts of the \bvfreq, respectively.
In the above formulae, $P$ is the pressure, $\chi_T \equiv (\partial \ln P/\partial \ln T)_\rho $, $\chi_\rho \equiv 
 (\partial \ln P/\partial \ln \rho)_T$, $\nabla_{\rm ad} = (\partial \ln T/\partial \ln P)_{S}$, and 
 $\nabla = d \ln T/d \ln P$ is the actual temperature gradient in the model. The term $B$ takes into 
 account composition gradients  and is defined as
\begin{equation}
 B = -\frac{1}{\chi_T} \frac{
   \ln P(\rho_k,T_k,\vec{X}_{k+1}) -
   \ln P(\rho_k,T_k,\vec{X}_{k})}{\ln P_{k+1} - \ln P_k},
 \label{bterm}
\end{equation}
where $\vec{X}_k$ is the vector of mass fractions of the different chemical species defined at the 
$k^{\rm th}$ mesh point. A neutrally buoyant profile ($N^2 \simeq 0$) will be one 
for which 
$B = \nabla - \nabla_{\rm ad}$;
this specifies the required gradient in the carbon (and oxygen) profile.

\begin{figure}[t!]
  \centering{\includegraphics[width=0.97\columnwidth]{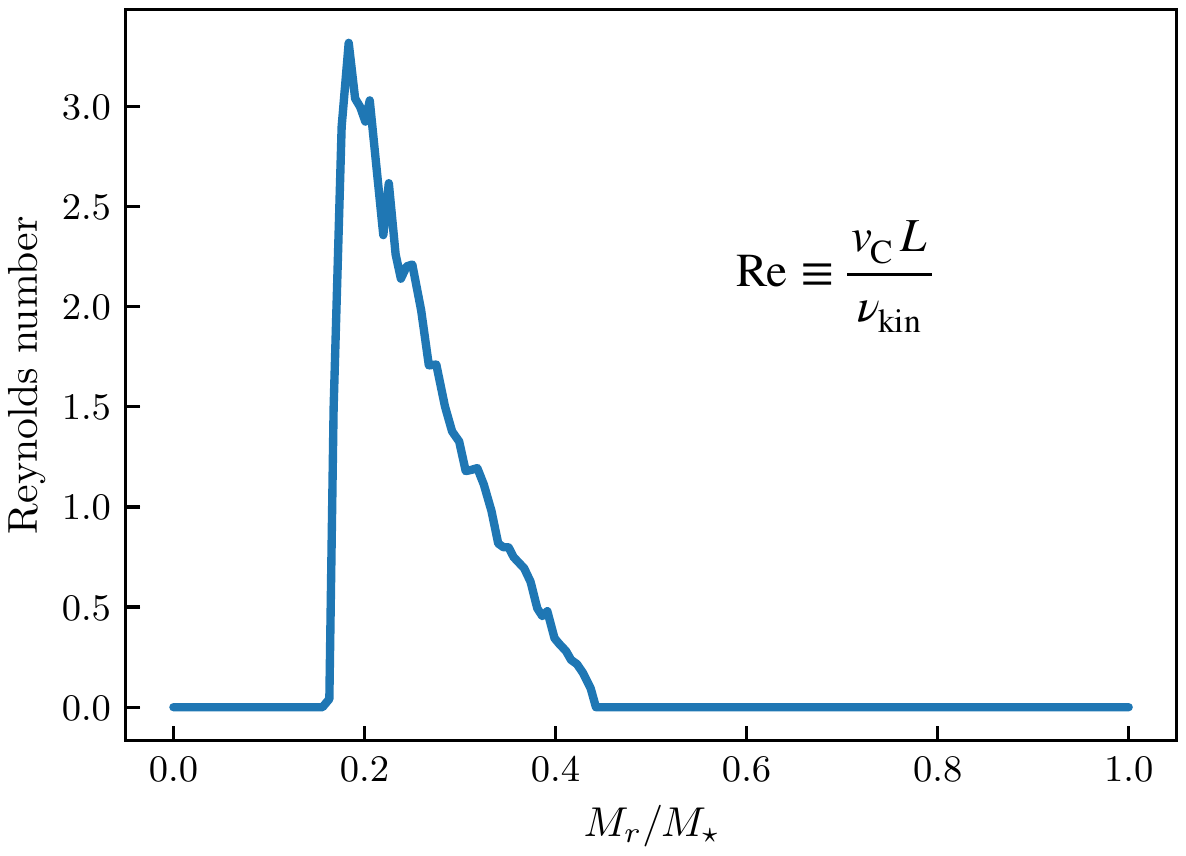}}
    \caption{The derived Reynolds number of the mixing for the case
    shown in Figure~\ref{xcomp}b. 
    We note that our calculation of 
    $\rm Re$ is independent of the mixing length $L$.
    }
  \label{Re_plot}
\end{figure}

We show the result of neutrally buoyant mixing in Figure~\ref{xcomp}a. For the sequence of five models shown, the mixed region now has a  positive oxygen gradient such that $N^2=0$. Also of note, unlike the uniform mixing case in which the mixing region extends out to  $M_r/M_\star \approx 0.9$ even for small amounts of crystallization, here the mixed region is smaller, only reaching the edge of the core  as the  crystallized mass fraction itself moves outward. 

In Figure~\ref{xcomp}b, we show the carbon flux required to produce this flux. 
Independent of any particular mixing length theory, the carbon enhancement of a fluid bubble,
$\delta X_C$, will be the difference in composition between the region in which it forms and the region in 
which it eventually mixes. Thus, if the fluid element travels a distance $L$, this enhancement is related 
to the C profile by
\begin{equation}
 \delta X_{\rm C} = \left|\frac{d X_{\rm C}}{d r}\right|\,L,
 \label{C_grad0}
\end{equation}
yielding, with equation~\ref{feqn}, the following equation for the flux:
\begin{equation}
  F_{\rm C} = f\,\left|\frac{dX_{\rm C}}{dr}\right|\, \rho \, L\, v_{\rm C}
  \label{feqn2}
\end{equation}
\citep[for an essentially identical result, see equation~19 of][]{Mochkovitch83}.

Within a factor of order unity, the Reynolds number, $\rm Re$, of this flow is approximately given by 
${\rm Re} = v_{\rm C}\, L/\nu_{\rm kin}$, where the kinematic viscosity, $\nu_{\rm kin}$, is approximately $0.4\,\rm cm^2\,s^{-1}$ for
these conditions \citep{Nandkumar84}. In Figure~\ref{Re_plot}, we show the value of $\rm Re$ for this
model. 
We see that $\rm Re$ is of order unity, which in general is much too small for the flow to be turbulent; typically, $\rm Re > 10^3$ is  required for turbulent flow.\footnote{For reference, water flow in a pipe becomes turbulent  for $\rm Re > 2300$ \citep[e.g.,][]{Menon15}.} While the idea of  ``laminar mixing'' may sound like a contradiction of terms, 
we note that laminar motions have been considered as part of the process for mixing heterogeneities in the Earth's mantle  \citep[e.g.,][]{Olson84a}.

\begin{figure}[t!]
  \centering{\includegraphics[width=0.98\columnwidth]{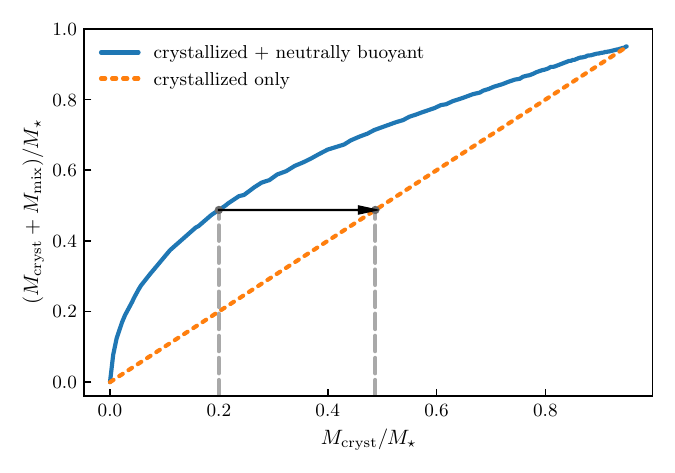}}
    \caption{The fractional mass of the neutrally mixed plus crystallized region
    (blue curve) as a function of the crystallized mass fraction. 
    The arrow indicates how a  model that is 20\% crystallized would have a crystallized 
    plus neutrally buoyant region making it appear $49$\% crystallized in models that assume uniform mixing.
    }
  \label{mix_map}
\end{figure}

\section{Pulsations and Neutral Profiles}
\label{pulse}

\citet{Montgomery99a} were the first to treat in detail the effect that a crystallized
core would have on the \gmode\ pulsations of WD stars. They found that \gmode s are
excluded from the crystallized solid region of the models. Due to this shrinking of the
\gmode\ cavity, the average period spacing of consecutive radial overtones of a given
$\ell$, $\langle\Delta P\rangle_\ell$, was found to be a monotonically increasing function
of the crystallized mass fraction. For instance, in going from uncrystallized to 90\%
crystallized by mass, while otherwise holding the thermal and mechanical structure
constant, $\langle\Delta P\rangle_{\ell=1}$ increases from 31~s to 43~s.

Since $g$ modes locally propagate in regions in which their squared frequency is
less than $N^2$, they will be excluded from neutrally mixed regions with $N^2 = 0$
as well as the crystallized core. In this case, we expect the effect of the mixed plus 
crystallized region to mimic that
of a star with a larger crystallized core. As we see in Figure~\ref{mix_map}, 
when our fiducial model is 20\% crystallized by mass it will have a crystallized plus 
mixed region that is almost 50\% of the total mass. Thus,
if we fit the pulsation frequencies of this model with a model that does not
take into account neutrally buoyant mixing, the fit will erroneously favor models that are
50\% crystallized by mass.
In fact, one prediction of this model is that stars analyzed with traditional seismic 
fits will be found to be either 0\% crystallized or $\gtrsim 40$\% crystallized by mass because the rapid growth of the neutrally buoyant region will make it unlikely to find 
them inside of these limits.
These general considerations should apply to all models which are 
massive enough to be partially crystallized while in the instability strip
\citep[e.g.,][]{Montgomery99a,Corsico05a,DeGeronimo19}.

In the following section, we examine the potential effects of thermohaline mixing on the chemical profiles.
If the thermohaline diffusion coefficient, $D_{\rm thrm}$, is sufficiently large, the resulting profiles will not be as steep as in the neutrally buoyant case.
The associated pulsational models will, therefore, also be different from those considered here.
However, the simulations used to derive $D_{\rm thrm}$ 
\citep[e.g.,][]{Brown13} must be extrapolated by $\sim$\,5 orders of magnitude in $\tau \equiv \kappa_{\mu}/\kappa_T$ (the ratio of molecular to thermal diffusivity) and 3 to 4 orders of magnitude in $\nu_{\rm kin}/\kappa_{\mu}$ (i.e., the Schmidt number) to match the WD interior conditions considered here (see Table~\ref{quants} for definitions and values of
these variables). Given this mismatch, neutral profiles cannot be definitively ruled out. 
In addition, asteroseismology could, in principle, provide evidence for the efficiency of the thermohaline mixing discussed below.

\section{Thermohaline Mixing}
\label{thermo}

The treatment of mixing in section~\ref{re-mix}
assumes that the  fluid elements do not exchange heat with their environment, i.e., the motion 
is adiabatic. In this case, fluid motions occur only when the chemical gradients are large enough 
that  $N^2 < 0$. However, when heat exchange with the environment is considered, smaller chemical
gradients can produce mixing in a process termed thermohaline convection or thermohaline
mixing. If this is sufficiently efficient, it could provide the required mixing during 
phase separation 
and prevent the formation of steep chemical gradients that would produce an RT instability.

In this section, we use MESA's thermohaline capabilities to see if the predicted fluxes of 
C and O due to thermohaline mixing are large enough to fulfill the mixing requirements of 
our models. 
In order to perform this calculation, we have modified MESA so that its thermohaline
prescription uses values for the kinematic viscosity $\nu_{\rm kin}$ \citep{Nandkumar84} 
and the molecular diffusivity  of the ions $\kappa_\mu$ \citep{Caplan21b, Caplan22} that are 
appropriate for WD stars. For these models we find 
$\kappa_\mu \approx 10^{-5}\rm \,  cm^2\,s^{-1}$ and $\nu_{\rm kin} \approx 0.4 \rm \,  cm^2\,s^{-1}$
(for these and other values, see Table~\ref{quants} in Appendix~\ref{thermohal}).

\begin{figure}[t!]
  \centering{\includegraphics[width=1.00\columnwidth]{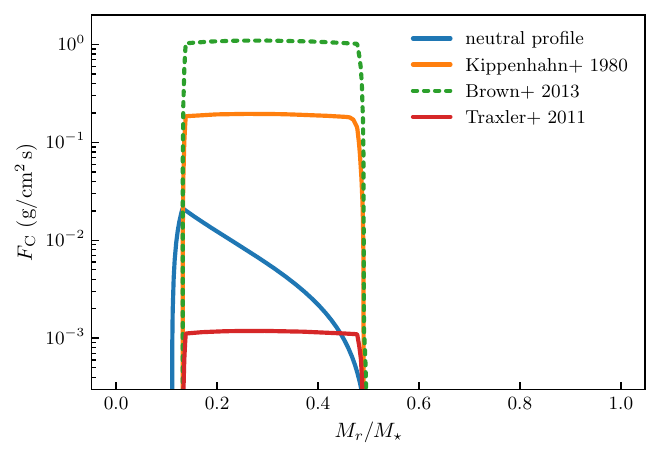}}
    \caption{ The flux of C due to different prescriptions for thermohaline mixing compared with the flux required to maintain a chemical profile that is neutrally buoyant (blue curve).
    We see that the \citet{Kippenhahn80} and \citet{Brown13} prescriptions can easily accommodate the 
    carbon flux due to phase separation. Only the \citet{Traxler11} prescription would be too inefficient
    to produce the required fluxes, presumably leading to the formation of neutrally buoyant profiles.
    }
  \label{thermo_h}
\end{figure}

To isolate this mixing from convective mixing, which would arise if at any point 
during convergence the profiles were to become too steep, we 
construct a chemical profile that is only 90\% as steep as a neutrally buoyant 
profile, i.e., a profile with $B = 0.9 \left( \nabla - \nabla_{\rm ad} \right)$, since
we can study this state in the absence of convection. Then, in 
the following time step, we turn on thermohaline mixing and observe the flux of carbon that 
this mixing produces. At this stage we have to be careful to reduce the time step
so that the profile is not appreciably altered during the time step. 
Since the flux scales as the chemical gradient to the $-1.62$ power ($R_0^{-1.62}$, see equation~\ref{fprop} in 
section~\ref{thermo_Brown}),  we divide the flux by $0.9^{1.62} \approx 0.84$ to 
estimate the flux that a neutrally buoyant profile would produce. 
The results of this procedure are plotted in Figure~\ref{thermo_h}. The blue curve 
shows the flux of C 
that  is required to maintain this profile as computed from two 
consecutive time steps. The orange curve shows the net flux of C due to thermohaline 
mixing that this profile would produce when using the formalism of \citet{Kippenhahn80}, 
while the green and red curves show the predicted fluxes when using the \citet{Brown13} and 
\citet{Traxler11} prescriptions, 
respectively. Since the fluxes due to the \citet{Kippenhahn80} and \citet{Brown13}
prescriptions exceed that needed to produce and maintain a neutrally buoyant profile, 
for these theories thermohaline mixing would be efficient enough to produce all the 
required mixing and the RT instability would not come in to play. On the other hand, the 
\citet{Traxler11} prescription produces mixing which is too weak to generate the required fluxes;
in this case, we suggest that a neutrally buoyant profile would form and the RT 
instability described  in the section~\ref{re-mix} would operate.

\section{A New Mixing Length Theory for Thermohaline Mixing}
\label{thermo2}

Given that the \citet{Brown13} prescription is the most recent, is derived from an underlying
physical model, and has been benchmarked against numerical simulations, we treat it as the 
currently favored theory for thermohaline processes. From Figure~\ref{thermo_h}, we see that it 
can transport a factor of 40 or more carbon than the models require, making it the dominant 
process for chemical transport during C/O phase separation. While we have not self-consistently 
computed the resulting chemical gradients, we can estimate that they will be reduced by a
factor of $40^{1/1.62} \approx 10$ from the neutrally buoyant case. Thus, the profiles
will be closer to the flat, uniformly mixed profiles of the standard case than the neutrally 
buoyant case.

On the other hand, \citet{Brown13} find somewhat less good agreement between their theory and the
simulations for the astrophysically interesting low-Pr, low-$\tau$ regime. This led us to 
re-examine their model, which we present in Appendix~\ref{thermo_Brown}.
Then, in Appendix~\ref{thermo_new}, we show that its general form can be derived as a mixing length 
theory and we suggest a small addition that accounts for the loss of compositional flux due to 
turbulent diffusivity. This decrease in efficiency of the chemical transport allows us to provide a closer 
match to their low-diffusivity, low-viscosity numerical simulations (see Figures~\ref{nusselt2}a,b).

Our new predicted fluxes are about a  factor of two smaller than those of \citet{Brown13}, and this is not large
enough to qualitatively change our conclusions in section~\ref{thermo}.
We estimate that the resulting chemical profiles should produce a
16\% reduction in $N^2$ compared to the uniform mixing case, a signature
with potential asteroseismological consequences.

\section{Magnetic Field Generation}

The high-mass overdensity defining the Q-branch contains an enhancement in the incidence of magnetic white dwarfs \citep{McCleery20}.  Among lower-mass WDs ($M_\star \la 0.75 \, M_\odot$), \citet{Bagnulo21,Bagnulo22} find an increasing incidence of
magnetic fields as they cool, particularly after the point when models of these
stars predict the onset of crystallization. In addition,
\citet{Parsons21} note a lack of identifiable progenitor systems for
magnetic cataclysmic variables. 
This leads \citet{Schreiber21,Schreiber22}
to speculate that the phenomenon of crystallization, which occurs at
late times, may be crucial for the generation of these fields. 
These groups cite \citet{Isern17} and/or \citet{Ginzburg22} as a possible explanation for this.
In a novel use of \citet{Isern17}, \citet{Camisassa22} infer the existence of O/Ne
cores based on the existence of magnetic fields in white dwarfs too hot to be crystallizing
if they have C/O cores, but which are expected to be crystallizing if they are O/Ne.

Based on work by \citet{Augustson16,Augustson19}, \citet{Ginzburg22}
use the following equation to estimate the strength of the dynamo-generated 
magnetic fields,
\begin{equation}
   B = \left( \frac{4 \pi \rho \, v_{\rm c}\, L}{P} \right)^{1/2},
   \label{bfield}
\end{equation}
where $P$ is the rotation period of the model. In their models they 
find $v_{\rm c}\,L \sim  10^{10}\,\rm cm^2 \,s^{-1}$. 
However, we show in section~\ref{inconsistency} that
their approach is not physically self-consistent, i.e., their assumption of rapid, adiabatic mixing is not consistent with an assumption of large mixing lengths.
In our neutrally buoyant mixing prescription, the carbon enhancement of the 
bubbles  is tied to the carbon gradient 
in the neutrally buoyant regions, so it is not possible in our model to have
both large mixing lengths and large velocities. Instead, we find 
$v_{\rm c}\,L \sim 1\,$cm$^2\,$s$^{-1}$. 
For the thermohaline case, we
find values that are slightly larger, with $v_{\rm c}\,L \sim 200\,$cm$^2\,$s$^{-1}$. 
Scaling from the results of \citet{Ginzburg22}  to our thermohaline results,
we predict magnetic fields 
$\sim 10^4$ times weaker, i.e., $B \la \rm 10 \,kG$. 
We conclude that the 
fluid mixing associated with phase separation is not a viable mechanism
for generating large magnetic fields in WDs.

However, this does not mean that crystallization itself cannot still 
play a role in magnetic field generation. If a WD is differentially rotating, 
either due to prior evolution or active accretion of angular momentum, then 
its rotational profile will be altered by crystallization. The crystallized core 
will rotate as a solid body while the overlying fluid layers will not. 
Thus, a strong shear layer could be set up at the solid/fluid boundary, and 
the fluid motions associated with this region plus the rotation of the WD
could lead to magnetic field generation \citep[e.g,][]{Spruit02}.

\section{Discussion}

Given the long time scale for crystallization, we have re-considered the assumption of
other studies that the mixing due to phase separation is vigorous and turbulent.
If thermohaline mixing is too inefficient,
we assume that slow RT mixing will produce neutrally buoyant profiles having $N^2 =0$, 
a result found in recent
studies \citep{Brooks17,Schwab19}, rather than the uniformly mixed profiles that are 
normally assumed. Simple estimates show that $v_c\, L \sim 1\,\rm cm^2 \,s^{-1}$, with a Reynolds number of $\mathcal{O}(1)$.

On the other hand, 
current theories of thermohaline convection predict that regions with 
shallower profiles than neutrally buoyant are capable of sustaining 
thermohaline fluxes large enough to transport the carbon released
during phase separation.
Assuming the  \citet{Brown13} formalism extrapolates to the physical
regime considered here, we find that $v_c\, L \sim 200\,\rm cm^2 \,s^{-1}$
and a Reynolds number of $\sim 500$. The theory also allows us to 
independently estimate the mixing length as the approximate 
length of the thermohaline ``fingers'' ($\sim 100$~cm) and the velocities as $\sim 2\,$cm$\,$s$^{-1}$.

In either the neutrally buoyant or thermohaline mixing case, we find that these
mixing lengths and velocities are much smaller than the estimates in \citet{Ginzburg22} and
\citet{Isern17}. In fact, the velocities and vigorous turbulent convection of those models 
result in physical inconsistencies; we thus conclude that such mixing is not a viable mechanism 
for generating large magnetic fields in WDs.

While we have confined our discussion to WDs with carbon/oxygen cores, everything 
we have  presented should apply equally well to WDs with crystallizing O/Ne/Mg cores. 
First, for the reasons cited in the above paragraphs, the fluid motions will not
be vigorous enough to produce measurable magnetic fields in these WDs. Second, 
for the neutrally buoyant mixing case,  the mapping of 
$M_{\rm cryst}$  in the standard treatment to $M_{\rm cryst} + M_{\rm mix}$ 
(as shown in Figure~\ref{mix_map}) should still hold. Although the cores of these
models may be mostly crystallized by the time the stars reach the DAV instability
strip, for high-mass DBs our results could be important for 
interpreting their pulsations as they begin to crystallize in the DBV instability strip.

Finally, we anticipate that implementing a thermohaline mixing prescription may be relevant for other
cases involving phase separation and mixing, such as the process of $^{22}\rm Ne$ 
distillation in ultramassive WDs as proposed by \citet{Blouin21a}. 

\vspace*{3em}

The authors thank Dean Townsley and Evan Bauer for useful discussions.
M.H.M. and B.H.D. acknowledge support from the Wootton Center for Astrophysical Plasma
Properties, a U.S. Department of Energy NNSA Stewardship Science Academic Alliance Center of Excellence supported under award numbers
DE-NA0003843 and DE-NA0004149, from the United States Department of Energy under grant DE-SC0010623,
and from the National Science Foundation under grant No.\ AST 1707419.
M.H.M. acknowledges support from the NASA ADAP program under grant 80NSSC20K0455.

\appendix
\vspace*{-2em}

\section{Numerical Implementation of Neutrally-Buoyant Mixing}
\label{num}

We implement this new mixing prescription in the stellar evolution
code \texttt{MESA}
\citep{Paxton11,Paxton13,Paxton15,Paxton18,Paxton19}. MESA provides
customization for different physical processes
through the \verb+run_star_extras.f+ file, which allows for a large
range of user-specified modifications without the need to modify any
central routines. The mixing is applied at the end of each time step in
the function \verb+extras_finish_step+, and heating due to the release
of gravitational energy is then applied during the next time step
through the \verb+other_energy+ hook in \verb+extras_controls+. These
calculations were performed using versions 12778 and 15140 of MESA.

We calculate $N^2$ and $B$ as given in equations~(\ref{bv0}) and
(\ref{bterm}) by calling \verb+do_brunt_N2+ and \verb+do_brunt_B+ in
MESA's \verb+brunt+ module.  We adjust the size of the mixed region
based on the amount of carbon that is liberated from the newly
crystallized shells.  While our procedure does not result in an $N^2$ 
which is identically  zero in these regions, it does reduce its value 
by at least a factor of $10^4$. This is sufficient both for computing
the energy release due to phase separation and for computing the effect
of a neutrally buoyant region on $g$-mode oscillations. 
We note that we
have also implemented the standard mixing prescription for phase
separation which produces uniform profiles in the mixed fluid region.

There is a final simplification we make. Crystallization and phase
separation occur in the highly degenerate interiors of WDs, and this
degeneracy leads to a decoupling of the thermal and hydrostatic
structure. We exploit this as follows. When the model first reaches
the temperature at which it begins to crystallize, we tabulate in
advance the chemical profile for an arbitrary crystallized mass
fraction ranging from 0\% to $\sim$100\% of the carbon/oxygen core, 
taking into account mixing
of the fluid layers according to either the standard prescription or
our proposed new prescription. We then compute the
gravitational/internal energy released by this process, again for an
arbitrary value of the crystallized mass fraction. While we do not
take into account the effect of the subsequent thermal evolution of
the model on the energy release, the error this introduces is quite
small. In \citet{Montgomery99}, we showed that this resulted in an
error of less than 0.5\% in the calculation of the total energy
released during phase separation.

With the composition profile and energy release due to phase separation
now tabulated as a function of crystallized mass fraction, we simply
interpolate at each subsequent time step to find the appropriate values
to use for the evolving WD model.

\section{Thermohaline Mixing}
\label{thermo_MLT}

In this section we suggest a possible addition to the \citet{Brown13} formalism. 
As a starting point, we first present their approach in section~\ref{thermo_Brown} below, followed 
by our modification in section~\ref{thermo_new}.

\subsection{The Formalism of Brown et al.}
\label{thermo_Brown}

The \citet{Brown13} prescription for thermohaline mixing represents the current state-of-the-art, 
particularly due to its ability to match the results of numerical simulations over a wide range 
of parameter  values, although these values still do not reach the regimes of most astrophysical 
objects. These simulations show an  initial instability evolving into a saturated, turbulent regime.
Since these simulations cannot currently be run for the parameter values found in actual stars, 
we need a theory or fitting formula that allows us to extrapolate to the cases of interest.

Using the Boussinesq approximation \citep{Spiegel60}, \citet{Brown13} and \citet{Garaud15} 
de-dimensionalize the fluid equations, with length given in units of $d = \left(\kappa_T \nu_{\rm kin}/ 
N_T^2\right)^{1/4}$ and time in units of  $d^2/\kappa_T = {\rm Pr}^{1/2}/N_T$. This yields
the following set of equations:
\begin{eqnarray}
    \nabla \cdot \vec{u} & = & 0 \\
    \frac{1}{\rm Pr} \left( \frac{\partial \vec{u}}{\partial t} + \vec{u} \cdot \nabla \vec{u} \right) 
    & = & \nabla P + (T - \mu) \vec{e_z} + \nabla^2 \vec{u} \\
    \frac{\partial T}{\partial t} + \vec{u} \cdot \nabla T + w & = & \nabla^2 T \\
    \frac{\partial \mu}{\partial t} + \vec{u} \cdot \nabla \mu + \frac{w}{R_0} & = & \tau \nabla^2 \mu.
    \label{mu}
\end{eqnarray}
Here, $P, T$, and $\mu$ are the perturbations of the pressure, temperature, and composition on top of
an assumed linearly-varying background state, $\vec{u}$ and $w$ are the total fluid velocity and its 
vertical component, respectively, $\rm Pr \equiv \nu_{\rm kin}/\kappa_T$, the Prandtl number, is the ratio
of the kinetic viscosity to the radiative diffusivity,
and $\tau \equiv \kappa_C/\kappa_T$ is the ratio of diffusivities of the composition to the temperature.
$R_0$ controls the ``strength'' of the instability, and is defined as 
\begin{equation}
    R_0 \equiv \frac{N_T^2}{|N_\mu^2|} = \frac{\nabla_{\rm ad} - \nabla}{|B|},
\end{equation}
where we have used terms defined in equations~\ref{bv0} and \ref{bv1}.

Next, \citet{Brown13} search for linearly growing solutions of the form
\begin{equation}
    \left(\vec{u}, w, T, \mu \right) = 
    \left(\vec{u}_0, w_0, T_0, \mu_0 \right)\,e^{i \vec{l} \cdot \vec{x} + \lambda t},
    \label{ansatz}
\end{equation}
where the quantities with 0 subscripts are constants, $\vec{l}$ is wavenumber in the $x$-$y$ plane,
and $\lambda$ is the growth rate of the instability. This leads to an eigenvalue problem
with $\lambda$ given by the solution of a cubic equation \citep[see also][]{Baines69}:
\begin{equation}
    \lambda^3 + a_2 \lambda^2 + a_1 \lambda + a_0 = 0, \mbox{\hspace*{1em} where}
    \label{cubic}
\end{equation}
\vspace*{-1.7em}
\begin{eqnarray}
    a_2 & = & l^2 (1 + {\rm Pr} + \tau), \\
    a_1 & = & l^4 (\tau {\rm Pr} + {\rm Pr} + \tau) + {\rm Pr}\left(1-R_0^{-1}\right), \\
    a_0 & = & l^6 \tau {\rm Pr} + l^2 {\rm Pr}\left(\tau - R_0^{-1}\right).
    \label{acoeffs}
\end{eqnarray}
To find the maximum growth rate, $\lambda_{\rm m}$, and its associated wavenumber, $l_{\rm m}$ 
they derive a new equation by taking a 
derivative of equation~\ref{cubic} with respect to $l$ and setting $d \lambda/d l =0$. They then
simultaneously solve these two equations for $\lambda_{\rm m}$ and $l_{\rm m}$.

\citet{Brown13} report their results as Nusselt numbers, which are defined as the total flux
divided by microscopic diffusive flux:
\begin{equation}
    {\rm Nu_{\mu}} = 1 - \frac{R_0}{\tau} \langle w \mu \rangle.
    \label{nuss_form}
\end{equation}
Next, to find the amplitudes of $w$ and $\mu$, they set the growth rate of a secondary shearing 
instability equal to that of the primary instability, arriving at the following expression:
\begin{equation}
    {\rm Nu_{\mu}} = 1 + C^2 \frac{\lambda^2}{\tau l^2 (\lambda + \tau l^2)}.
\end{equation}
Thus, $\rm Nu_{\mu}$ is a function of Pr, $\tau$, and $R_0$. We note that assuming $R_0 \sim O(1)$
and solving the above equations in the limit $\tau \ll {\rm Pr} \ll 1$ yields 
\begin{equation}
l \approx 1, \hspace{2em}
\lambda_{\rm m} \approx \left(\frac{\rm Pr}{R_0}\right)^{1/2}, \hspace{2em}
{\rm Nu}_{\mu} - 1  \approx \frac{C^2}{\tau}\, \lambda = 
         \frac{C^2}{\tau}\, \left(\frac{\rm Pr}{R_0}\right)^{1/2}
    \label{asymp}
\end{equation}
In Figure~\ref{Rplot}, we plot $D_{\rm thrm} \propto \rm Nu_\mu - 1$ as a function of $R_0$ 
for the values of $\tau$ and Pr given in Table~\ref{quants}.
While close to $R_0^{-1/2}$, we find the actual scaling to be $D_{\rm thrm} \propto
R_0^{-0.62}$. Since the flux $F_\mu$ is proportional to ${\rm Nu}_\mu$ times the $\mu$ 
gradient, which in turn is proportional to 1/$R_0$, we find that for our case the flux 
should scale as
\begin{equation}
    F_\mu \propto R_0^{-1.62}.
    \label{fprop}
\end{equation}

To aid in comparing and plotting the different simulations they remap the ``fingering regime'' of 
$R_0$, $1 < R_0 < \tau^{-1}$, onto the unit interval, defining
\begin{equation}
    r = \frac{R_0 -1}{\tau^{-1}-1}.
    \label{rdef}
\end{equation}
From fits to their numerical simulations, they find $C \approx 7$ and are able to 
provide a remarkably good fit over a wide range of parameters (see Figure~\ref{nusselt2}a). 
The predicted horizontal size of their ``bubbles'' is $2 \pi/l_{\rm m} \approx 10\,d$, which
also matches the simulations well. On the other hand, for the
lowest Pr and $\tau$ values, there is about a factor of two mis-match between the theoretical
and numerically simulated value of Nu$_\mu$, and these values are the closest to the regimes 
of astrophysical interest.

\subsection{A New MLT Derivation}
\label{thermo_new}

In addition to equations~\ref{cubic} -- \ref{acoeffs} for the eigenvalue $\lambda$, we also need
relations between the amplitudes of the various perturbations on the RHS of equation~\ref{ansatz}.
In particular, from equation~26 of \citet{Brown13} we have
\begin{equation}
    \mu_0 = - \frac{R_0^{-1}}{\lambda + \tau l^2} \,w_0.
    \label{mu2}
\end{equation}
Substituting 
\begin{equation}
    \langle \mu w \rangle \approx \mu_0 w_0
\end{equation}
in equation~\ref{nuss_form} yields
\begin{equation}
    {\rm Nu_{\mu}} = 1 + \frac{w_0^2}{\tau (\lambda + \tau l^2)}.
\end{equation}
To estimate the velocity, $w_0$, we take the growth rate, $\lambda$, times the mixing length, $2 \pi/l$,
which yields
\begin{equation}
    {\rm Nu_{\mu}} = 1 + C^2 \frac{\lambda^2}{\tau l^2 (\lambda + \tau l^2)},
\end{equation}
where $C^2 = (2\pi)^2$.
We note that the only difference between this and the \citet{Brown13} formula is our prediction 
for the value of $C^2$.

We now wish to apply a correction that takes into account turbulent mixing. In a turbulent medium, a fluid
element will be continuously mixed with its environment, which will tend to reduce the vertical
flux. We can model this as a diffusive process, with an additional diffusion term. For example,
we can make the following replacement on the RHS of equation~\ref{mu}: 
\begin{equation}
    \tau \nabla^2 \mu \rightarrow \left( \tau + D \right) \nabla^2 \mu.
\end{equation}
Since $D$ is due to turbulent motions, we estimate it as a velocity times a length scale, i.e.,
$D \sim \psi \, \lambda/l^2$, where $\psi$ is a dimensionless constant we will later adjust.

Since
\begin{equation}
    D \,\nabla^2 \mu \simeq 
    \left(\frac{\psi \lambda}{l^2}\right) \left(l^2 \mu \right)
    = \psi\,\lambda\,\mu,
\end{equation}
the addition of a such a term in equation~\ref{mu} modifies equation~\ref{mu2} so that it becomes
\begin{equation}
    \mu_0 = - \frac{R_0^{-1}}{\lambda (1+\psi) + \tau l^2} \,w_0.
\end{equation}
Thus, the only modification is that
\begin{equation}
    \lambda \rightarrow \lambda (1 + \psi).
\end{equation}
Adding a term to the right-hand sides of the other equations also results in $\lambda$ being replaced
by $\lambda (1 + \psi)$. If we define $\lambda_0 \equiv \lambda (1+\psi)$ as the solution of the 
unmodified eigenvalue 
equation~\ref{cubic}, then we find
\begin{eqnarray}
    {\rm Nu_{\mu}} & = & 1 + C^2 \frac{\lambda^2}{\tau l^2 (\lambda_0 + \tau l^2)} \\
    & = & 1 + C^2 \frac{\lambda_0^2}{\tau l^2 (1+\psi)^2 (\lambda_0 + \tau l^2)}.
    \label{Nu_new}
\end{eqnarray}
We find that setting $\psi = 0.35$ in equation~\ref{Nu_new} results in a good match to the 
low-Pr, low-$\tau$ results. On the other hand, the match to many of the other results is 
noticeably worse, and the overall fit to the data is not as good as that of \citet{Brown13}.

We posit that the above correction should only be applied when the fluid flow is actually turbulent, which 
should be when the Reynolds number, Re, exceeds some critical value. We use the following ad hoc prescription 
to calculate $\psi$:
\begin{equation}
    \psi = 
\begin{cases}
    0,     & \text{if } {\rm Re} < \eta\\
    0.35,  & \text{if } {\rm Re} \geq \eta,
\end{cases}
\end{equation}
where
\vspace*{-1em}
\begin{equation}
    {\rm Re} = (2 \pi)^2 \frac{\lambda}{{\rm Pr}\, l_{\rm m}^2} \mbox{\hspace*{1em} and\hspace*{1em}} \eta = 2000 \, {\rm Pr}.
\end{equation}
We provide no justification for the Pr dependence of the threshold value $\eta$ other than it
better matches the data.

\begin{figure*}[t!]
  \includegraphics[width=0.49\columnwidth]{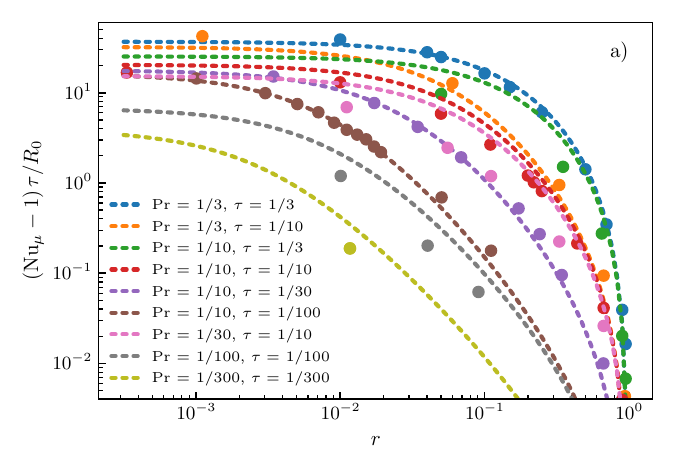}
  \includegraphics[width=0.49\columnwidth]{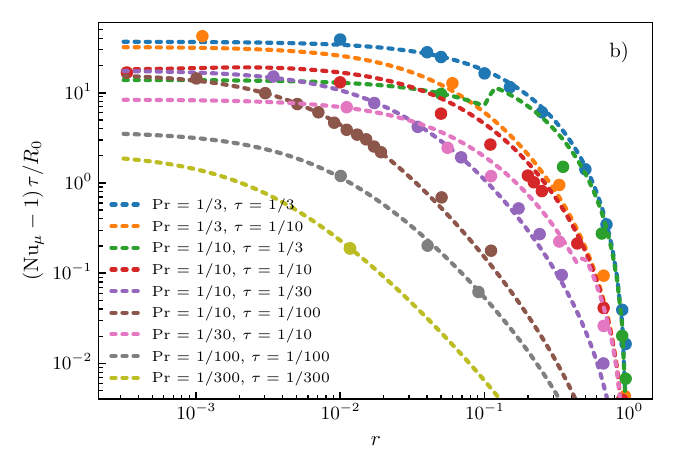}
    \caption{
    (a) The Nusselt number (filled circles) as calculated in the simulations of \citet{Brown13} 
    together with their theoretical model (dotted lines).
    (b) The same as (a) but using our new prescription.
    Our modified expression allows us to provide a better match to the low Pr and $\tau$ 
    simulations, which are closer to the astrophysical regimes of interest.
    }
  \label{nusselt2}
\end{figure*}

Figure~\ref{nusselt2}a shows the original fits of \citet{Brown13} to their simulation data while
Figure~\ref{nusselt2}b shows our new fits to these same data. For our fits we also take $C = 7$, and we
choose the value of $\psi = 0.35$ to match the low-Pr,  low-$\tau$ results. The threshold 
$\eta = 2000 \, \rm Pr$ was chosen to improve the fits for other cases. For instance, this resulted 
in an improved fit to the ${\rm Pr} = 1/30$, $\tau = 1/10$ case, which is poorly fit by
\citet{Brown13}. Of course, our fits \emph{should} improve since we are increasing the
number of fit parameters from 1 to 3.

We note a recent paper by \citet{Fuentes23} presents a mixing length theory for 
chemical transport that treats in a unified way both the thermohaline and  convective
regimes. A significant difference with our present work is that they take the mixing
length to be of order a pressure scale height, whereas we use a much smaller value of 
$\sim 10 d \approx 100\,$cm.
While they have made a careful treatment of heat exchange and other
diffusive processes, it is not yet clear how their work relates to formal 
treatments of thermohaline mixing, such as that of \citet{Brown13}.

\subsection{Spanning the $R_0 > 1$ and $R_0 < 1$ Regimes}

\begin{figure*}[t!]
  \centering{
  \includegraphics[width=0.70\columnwidth]{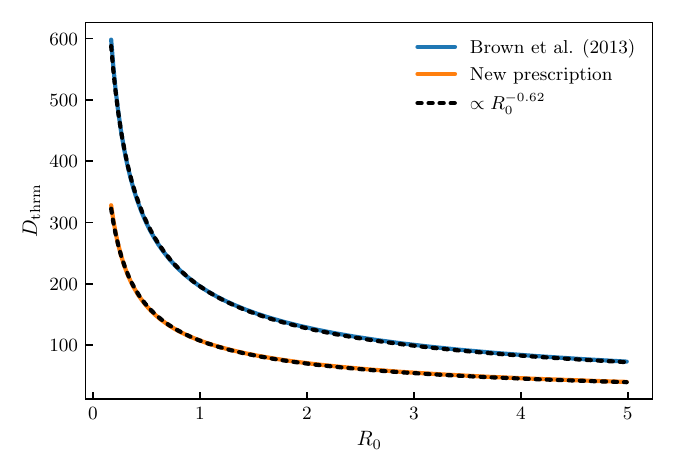}
  }
    \caption{
    The thermohaline mixing coefficient (in cgs units) as a function of $R_0$.
    The region $R_0 > 1$ represents the standard thermohaline mixing case, while $R_0 < 1$
    is the RT convective region. We see that our new prescription has a mixing efficiency about half 
    as large as that of \citet{Brown13}, for the model WD considered here.}
  \label{Rplot}
\end{figure*}
    
If we consider $\lambda_m$ and ${\rm Nu}_\mu$ as a function $R_0$ (instead of $r$, see equation~\ref{rdef}), 
then we find that our MLT approach can naturally be extended to the Ledoux convective (RT) regime. 
In Figure~\ref{Rplot} we plot the thermohaline mixing coefficient (in cgs units) as a function of $R_0$.
We have chosen representative values for Pr, $\kappa_T$, and $\kappa_\mu$ as given in Table~\ref{quants}.
As advertised, our prescription is about half as efficient as that of \citet{Brown13}. Each dashed black
curve shows a $R_0^{-0.62}$ scaling of the value of $D_{\rm thrm}$ at $R_0=1$, 
which is different from but similar to the expected asymptotic scaling of $R_0^{-1/2}$ (equation~\ref{asymp}).

Applying this prescription in the RT regime means using $L = 2 \pi/l_{\rm m}$ as the 
mixing length. While $L$ does increase as $R_0$ decreases, it is still very much less than the 
traditional choice of a pressure scale height. It is quite possible, or even likely, that additional 
instabilities occur in this regime which lead the actual mixing length to be of order a 
pressure scale height. This would lead to an increase in mixing of several orders of magnitude. 
Even so, the values shown in Figure~\ref{Rplot} should provide a lower bound for the
mixing in this regime.
A numerical exploration of the transition from thermohaline to RT convection would help clarify
this transition.

\section{Parameter Values for the Thermohaline Case}
\label{thermohal}

Models of thermohaline mixing depend on the microphysical properties of the fluid in question.
These can be transport properties 
(e.g., viscosity and compositional and radiative diffusivities) 
or thermodynamic properties (e.g., $\chi_T$, $\chi_\rho$), and can involve the dimensionless
ratios of these quantities (e.g., Pr, $\tau$).

\begin{deluxetable}{clc}
\label{quants}
\tablecolumns{3}
\tablecaption{Table of Symbol Definitions and Typical Values\label{defs}}
\tablehead{
 \colhead{Symbol} & \colhead{Definition}
 & \colhead{Values in WD Model (cgs)} 
}
\startdata
$\nu_{\rm kin}$ & kinematic viscosity & 0.4 \\
$\kappa_T$ & thermal diffusivity & 40 \\
$\kappa_\mu$ & compositional diffusivity & $10^{-5}$ \\
$\chi_T$ & $= (\partial \ln P/\partial \ln T)_\rho$ & $6 \cdot 10^{-4}$ \\
$\chi_\rho$ & $= (\partial \ln P/\partial \ln \rho)_T$ & 1.5 \\
$\alpha$ & $= \chi_T/T \chi_\rho$ & $6 \cdot 10^{-11}$ \\
$g$ & acceleration due to gravity & $10^9$ \\
$d$ &  $= (\kappa_T \nu_{\rm kin}/g \alpha |T_{0z} - T_{0z}^{\rm ad}| )^{1/4}$ (expected finger width) & 10 \\
Pr & $= \nu_{\rm kin}/\kappa_T$  (Prandtl number) & 0.01 \\
$\tau$ & $= \kappa_\mu/\kappa_T$ & $3 \cdot 10^{-7}$\\
$D_{\rm th}$ & thermohaline diffusion coefficient \citep{Brown13} & 200 \\
$L$ & $= 10 d$ (typical finger size) & 100 \\
$v_{\rm th}$ & $= D_{\rm th}/ L$ & 2\\
Re &  $ = D_{\rm th}/\nu_{\rm kin}$ (Reynolds number) & 500 \\
\enddata
\end{deluxetable}

In Table~\ref{quants}, we list several quantities that are important for the discussion 
in this paper. We also list approximate values in cgs units for these quantities as obtained from
our 1$\,$\msun WD model.
In the degenerate interior of a WD,
the mean free path of the electrons becomes longer, and this leads to a much higher value for the
viscosity than is found in non-degenerate stars. 
We estimate a Reynolds number for the thermohaline mixing that is $\sim 500$
and a Prandtl number,
which  is the ratio of the kinetic viscosity and the thermal diffusivity, 
of order Pr~$\sim 10^{-2}$.

\bibliographystyle{aasjournal}
\bibliography{index_bibdesk}

\begin{thebibliography}{}
\expandafter\ifx\csname natexlab\endcsname\relax\def\natexlab#1{#1}\fi
\providecommand{\url}[1]{\href{#1}{#1}}
\providecommand{\dodoi}[1]{doi:~\href{http://doi.org/#1}{\nolinkurl{#1}}}
\providecommand{\doeprint}[1]{\href{http://ascl.net/#1}{\nolinkurl{http://ascl.net/#1}}}
\providecommand{\doarXiv}[1]{\href{https://arxiv.org/abs/#1}{\nolinkurl{https://arxiv.org/abs/#1}}}

\bibitem[{{Althaus} {et~al.}(2003){Althaus}, {Serenelli}, {C{\'o}rsico}, \&
  {Montgomery}}]{Althaus03}
{Althaus}, L.~G., {Serenelli}, A.~M., {C{\'o}rsico}, A.~H., \& {Montgomery},
  M.~H. 2003, \aap, 404, 593, \dodoi{10.1051/0004-6361:20030472}

\bibitem[{{Augustson} {et~al.}(2016){Augustson}, {Brun}, \&
  {Toomre}}]{Augustson16}
{Augustson}, K.~C., {Brun}, A.~S., \& {Toomre}, J. 2016, \apj, 829, 92,
  \dodoi{10.3847/0004-637X/829/2/92}

\bibitem[{{Augustson} {et~al.}(2019){Augustson}, {Brun}, \&
  {Toomre}}]{Augustson19}
---. 2019, \apj, 876, 83, \dodoi{10.3847/1538-4357/ab14ea}

\bibitem[{{Bagnulo} \& {Landstreet}(2021)}]{Bagnulo21}
{Bagnulo}, S., \& {Landstreet}, J.~D. 2021, \mnras, 507, 5902,
  \dodoi{10.1093/mnras/stab2046}

\bibitem[{{Bagnulo} \& {Landstreet}(2022)}]{Bagnulo22}
---. 2022, \apjl, 935, L12, \dodoi{10.3847/2041-8213/ac84d3}

\bibitem[{{Baines} \& {Gill}(1969)}]{Baines69}
{Baines}, P.~G., \& {Gill}, A.~E. 1969, Journal of Fluid Mechanics, 37, 289,
  \dodoi{10.1017/S0022112069000553}

\bibitem[{{Bauer}(2023)}]{Bauer23}
{Bauer}, E.~B. 2023, \apj, 950, 115, \dodoi{10.3847/1538-4357/acd057}

\bibitem[{{Bauer} {et~al.}(2020){Bauer}, {Schwab}, {Bildsten}, \&
  {Cheng}}]{Bauer20}
{Bauer}, E.~B., {Schwab}, J., {Bildsten}, L., \& {Cheng}, S. 2020, \apj, 902,
  93, \dodoi{10.3847/1538-4357/abb5a5}

\bibitem[{{Blouin} {et~al.}(2021){Blouin}, {Daligault}, \&
  {Saumon}}]{Blouin21a}
{Blouin}, S., {Daligault}, J., \& {Saumon}, D. 2021, \apjl, 911, L5,
  \dodoi{10.3847/2041-8213/abf14b}

\bibitem[{{Blouin} {et~al.}(2020){Blouin}, {Daligault}, {Saumon}, {B{\'e}dard},
  \& {Brassard}}]{Blouin20a}
{Blouin}, S., {Daligault}, J., {Saumon}, D., {B{\'e}dard}, A., \& {Brassard},
  P. 2020, \aap, 640, L11, \dodoi{10.1051/0004-6361/202038879}

\bibitem[{{Brooks} {et~al.}(2017){Brooks}, {Schwab}, {Bildsten}, {Quataert}, \&
  {Paxton}}]{Brooks17}
{Brooks}, J., {Schwab}, J., {Bildsten}, L., {Quataert}, E., \& {Paxton}, B.
  2017, \apjl, 834, L9, \dodoi{10.3847/2041-8213/834/2/L9}

\bibitem[{{Brown} {et~al.}(2013){Brown}, {Garaud}, \& {Stellmach}}]{Brown13}
{Brown}, J.~M., {Garaud}, P., \& {Stellmach}, S. 2013, \apj, 768, 34,
  \dodoi{10.1088/0004-637X/768/1/34}

\bibitem[{{Camisassa} {et~al.}(2022){Camisassa}, {Raddi}, {Althaus}, {Isern},
  {Rebassa-Mansergas}, {Torres}, {C{\'o}rsico}, \& {Korre}}]{Camisassa22}
{Camisassa}, M.~E., {Raddi}, R., {Althaus}, L.~G., {et~al.} 2022, \mnras, 516,
  L1, \dodoi{10.1093/mnrasl/slac078}

\bibitem[{{Caplan} {et~al.}(2022){Caplan}, {Bauer}, \& {Freeman}}]{Caplan22}
{Caplan}, M.~E., {Bauer}, E.~B., \& {Freeman}, I.~F. 2022, \mnras, 513, L52,
  \dodoi{10.1093/mnrasl/slac032}

\bibitem[{{Caplan} \& {Freeman}(2021)}]{Caplan21b}
{Caplan}, M.~E., \& {Freeman}, I.~F. 2021, \mnras, 505, 45,
  \dodoi{10.1093/mnras/stab1259}

\bibitem[{{Cheng} {et~al.}(2019){Cheng}, {Cummings}, \& {M{\'e}nard}}]{Cheng19}
{Cheng}, S., {Cummings}, J.~D., \& {M{\'e}nard}, B. 2019, \apj, 886, 100,
  \dodoi{10.3847/1538-4357/ab4989}

\bibitem[{{Christensen}(2010)}]{Christensen10}
{Christensen}, U.~R. 2010, \ssr, 152, 565, \dodoi{10.1007/s11214-009-9553-2}

\bibitem[{{C{\'o}rsico} {et~al.}(2005){C{\'o}rsico}, {Althaus}, {Montgomery},
  {Garc{\'{\i}}a-Berro}, \& {Isern}}]{Corsico05a}
{C{\'o}rsico}, A.~H., {Althaus}, L.~G., {Montgomery}, M.~H.,
  {Garc{\'{\i}}a-Berro}, E., \& {Isern}, J. 2005, \aap, 429, 277

\bibitem[{{De Ger{\'o}nimo} {et~al.}(2019){De Ger{\'o}nimo}, {C{\'o}rsico},
  {Althaus}, {Wachlin}, \& {Camisassa}}]{DeGeronimo19}
{De Ger{\'o}nimo}, F.~C., {C{\'o}rsico}, A.~H., {Althaus}, L.~G., {Wachlin},
  F.~C., \& {Camisassa}, M.~E. 2019, \aap, 621, A100,
  \dodoi{10.1051/0004-6361/201833789}

\bibitem[{{Fuentes} {et~al.}(2023){Fuentes}, {Cumming}, {Castro-Tapia}, \&
  {Anders}}]{Fuentes23}
{Fuentes}, J.~R., {Cumming}, A., {Castro-Tapia}, M., \& {Anders}, E.~H. 2023,
  arXiv e-prints, arXiv:2301.04273, \dodoi{10.48550/arXiv.2301.04273}

\bibitem[{{Garaud} {et~al.}(2015){Garaud}, {Medrano}, {Brown}, {Mankovich}, \&
  {Moore}}]{Garaud15}
{Garaud}, P., {Medrano}, M., {Brown}, J.~M., {Mankovich}, C., \& {Moore}, K.
  2015, \apj, 808, 89, \dodoi{10.1088/0004-637X/808/1/89}

\bibitem[{{Ginzburg} {et~al.}(2022){Ginzburg}, {Fuller}, {Kawka}, \&
  {Caiazzo}}]{Ginzburg22}
{Ginzburg}, S., {Fuller}, J., {Kawka}, A., \& {Caiazzo}, I. 2022, \mnras, 514,
  4111, \dodoi{10.1093/mnras/stac1363}

\bibitem[{{Horowitz} {et~al.}(2010){Horowitz}, {Schneider}, \&
  {Berry}}]{Horowitz10}
{Horowitz}, C.~J., {Schneider}, A.~S., \& {Berry}, D.~K. 2010, Physical Review
  Letters, 104, 231101, \dodoi{10.1103/PhysRevLett.104.231101}

\bibitem[{{Isern} {et~al.}(2017){Isern}, {Garc{\'{\i}}a-Berro}, {K{\"u}lebi},
  \& {Lor{\'e}n-Aguilar}}]{Isern17}
{Isern}, J., {Garc{\'{\i}}a-Berro}, E., {K{\"u}lebi}, B., \&
  {Lor{\'e}n-Aguilar}, P. 2017, \apjl, 836, L28,
  \dodoi{10.3847/2041-8213/aa5eae}

\bibitem[{{Isern} {et~al.}(1997){Isern}, {Mochkovitch}, {Garc\'{\i}a-Berro}, \&
  {Hernanz}}]{Isern97}
{Isern}, J., {Mochkovitch}, R., {Garc\'{\i}a-Berro}, E., \& {Hernanz}, M. 1997,
  \apj, 485, 308

\bibitem[{{Kippenhahn} {et~al.}(1980){Kippenhahn}, {Ruschenplatt}, \&
  {Thomas}}]{Kippenhahn80}
{Kippenhahn}, R., {Ruschenplatt}, G., \& {Thomas}, H.~C. 1980, \aap, 91, 175

\bibitem[{{McCleery} {et~al.}(2020){McCleery}, {Tremblay}, {Gentile Fusillo},
  {Hollands}, {G{\"a}nsicke}, {Izquierdo}, {Toonen}, {Cunningham}, \&
  {Rebassa-Mansergas}}]{McCleery20}
{McCleery}, J., {Tremblay}, P.-E., {Gentile Fusillo}, N.~P., {et~al.} 2020,
  \mnras, 499, 1890, \dodoi{10.1093/mnras/staa2030}

\bibitem[{Medin \& Cumming(2010)}]{Medin10}
Medin, Z., \& Cumming, A. 2010, Phys. Rev. E, 81, 036107,
  \dodoi{10.1103/PhysRevE.81.036107}

\bibitem[{{Mochkovitch}(1983)}]{Mochkovitch83}
{Mochkovitch}, R. 1983, \aap, 122, 212

\bibitem[{{Montgomery} {et~al.}(1999){Montgomery}, {Klumpe}, {Winget}, \&
  {Wood}}]{Montgomery99}
{Montgomery}, M.~H., {Klumpe}, E.~W., {Winget}, D.~E., \& {Wood}, M.~A. 1999,
  \apj, 525, 482, \dodoi{10.1086/307871}

\bibitem[{{Montgomery} \& {Winget}(1999)}]{Montgomery99a}
{Montgomery}, M.~H., \& {Winget}, D.~E. 1999, \apj, 526, 976,
  \dodoi{10.1086/308044}

\bibitem[{{Nandkumar} \& {Pethick}(1984)}]{Nandkumar84}
{Nandkumar}, R., \& {Pethick}, C.~J. 1984, \mnras, 209, 511,
  \dodoi{10.1093/mnras/209.3.511}

\bibitem[{{Olson} {et~al.}(1984){Olson}, {Yuen}, \& {Balsiger}}]{Olson84a}
{Olson}, P., {Yuen}, D.~A., \& {Balsiger}, D. 1984, \jgr, 89, 425,
  \dodoi{10.1029/JB089iB01p00425}

\bibitem[{{Parsons} {et~al.}(2021){Parsons}, {G{\"a}nsicke}, {Schreiber},
  {Marsh}, {Ashley}, {Breedt}, {Littlefair}, \& {Meusinger}}]{Parsons21}
{Parsons}, S.~G., {G{\"a}nsicke}, B.~T., {Schreiber}, M.~R., {et~al.} 2021,
  \mnras, 502, 4305, \dodoi{10.1093/mnras/stab284}

\bibitem[{{Paxton} {et~al.}(2011){Paxton}, {Bildsten}, {Dotter}, {Herwig},
  {Lesaffre}, \& {Timmes}}]{Paxton11}
{Paxton}, B., {Bildsten}, L., {Dotter}, A., {et~al.} 2011, \apjs, 192, 3,
  \dodoi{10.1088/0067-0049/192/1/3}

\bibitem[{{Paxton} {et~al.}(2013){Paxton}, {Cantiello}, {Arras}, {Bildsten},
  {Brown}, {Dotter}, {Mankovich}, {Montgomery}, {Stello}, {Timmes}, \&
  {Townsend}}]{Paxton13}
{Paxton}, B., {Cantiello}, M., {Arras}, P., {et~al.} 2013, \apjs, 208, 4,
  \dodoi{10.1088/0067-0049/208/1/4}

\bibitem[{{Paxton} {et~al.}(2015){Paxton}, {Marchant}, {Schwab}, {Bauer},
  {Bildsten}, {Cantiello}, {Dessart}, {Farmer}, {Hu}, {Langer}, {Townsend},
  {Townsley}, \& {Timmes}}]{Paxton15}
{Paxton}, B., {Marchant}, P., {Schwab}, J., {et~al.} 2015, \apjs, 220, 15,
  \dodoi{10.1088/0067-0049/220/1/15}

\bibitem[{{Paxton} {et~al.}(2018){Paxton}, {Schwab}, {Bauer}, {Bildsten},
  {Blinnikov}, {Duffell}, {Farmer}, {Goldberg}, {Marchant}, {Sorokina},
  {Thoul}, {Townsend}, \& {Timmes}}]{Paxton18}
{Paxton}, B., {Schwab}, J., {Bauer}, E.~B., {et~al.} 2018, \apjs, 234, 34,
  \dodoi{10.3847/1538-4365/aaa5a8}

\bibitem[{{Paxton} {et~al.}(2019){Paxton}, {Smolec}, {Schwab}, {Gautschy},
  {Bildsten}, {Cantiello}, {Dotter}, {Farmer}, {Goldberg}, {Jermyn}, {Kanbur},
  {Marchant}, {Thoul}, {Townsend}, {Wolf}, {Zhang}, \& {Timmes}}]{Paxton19}
{Paxton}, B., {Smolec}, R., {Schwab}, J., {et~al.} 2019, \apjs, 243, 10,
  \dodoi{10.3847/1538-4365/ab2241}

\bibitem[{{Salaris} {et~al.}(1997){Salaris}, {Dominguez}, {Garcia-Berro},
  {Hernanz}, {Isern}, \& {Mochkovitch}}]{Salaris97}
{Salaris}, M., {Dominguez}, I., {Garcia-Berro}, E., {et~al.} 1997, \apj, 486,
  413, \dodoi{10.1086/304483}

\bibitem[{{Schneider} {et~al.}(2012){Schneider}, {Hughto}, {Horowitz}, \&
  {Berry}}]{Schneider12}
{Schneider}, A.~S., {Hughto}, J., {Horowitz}, C.~J., \& {Berry}, D.~K. 2012,
  \pre, 85, 066405, \dodoi{10.1103/PhysRevE.85.066405}

\bibitem[{{Schreiber} {et~al.}(2021){Schreiber}, {Belloni}, {G{\"a}nsicke},
  {Parsons}, \& {Zorotovic}}]{Schreiber21}
{Schreiber}, M.~R., {Belloni}, D., {G{\"a}nsicke}, B.~T., {Parsons}, S.~G., \&
  {Zorotovic}, M. 2021, Nature Astronomy, 5, 648,
  \dodoi{10.1038/s41550-021-01346-8}

\bibitem[{{Schreiber} {et~al.}(2022){Schreiber}, {Belloni}, {Zorotovic},
  {Zapata}, {G{\"a}nsicke}, \& {Parsons}}]{Schreiber22}
{Schreiber}, M.~R., {Belloni}, D., {Zorotovic}, M., {et~al.} 2022, \mnras, 513,
  3090, \dodoi{10.1093/mnras/stac1076}

\bibitem[{{Schwab} \& {Garaud}(2019)}]{Schwab19}
{Schwab}, J., \& {Garaud}, P. 2019, \apj, 876, 10,
  \dodoi{10.3847/1538-4357/ab113f}

\bibitem[{{Segretain} \& {Chabrier}(1993)}]{Segretain93}
{Segretain}, L., \& {Chabrier}, G. 1993, \aap, 271, L13

\bibitem[{{Shashi Menon}(2015)}]{Menon15}
{Shashi Menon}, E. 2015, in Transmission Pipeline Calculations and Simulations
  Manual, ed. E.~{Shashi Menon} (Boston: Gulf Professional Publishing),
  149--234, \dodoi{https://doi.org/10.1016/B978-1-85617-830-3.00005-5}

\bibitem[{{Spiegel} \& {Veronis}(1960)}]{Spiegel60}
{Spiegel}, E.~A., \& {Veronis}, G. 1960, \apj, 131, 442, \dodoi{10.1086/146849}

\bibitem[{{Spruit}(2002)}]{Spruit02}
{Spruit}, H.~C. 2002, \aap, 381, 923, \dodoi{10.1051/0004-6361:20011465}

\bibitem[{{Stevenson}(1980)}]{Stevenson80}
{Stevenson}, D.~J. 1980, Journal de Physique, 41, C2\_61

\bibitem[{{Traxler} {et~al.}(2011){Traxler}, {Garaud}, \&
  {Stellmach}}]{Traxler11}
{Traxler}, A., {Garaud}, P., \& {Stellmach}, S. 2011, \apjl, 728, L29,
  \dodoi{10.1088/2041-8205/728/2/L29}

\bibitem[{{Tremblay} {et~al.}(2019){Tremblay}, {Fontaine}, {Fusillo}, {Dunlap},
  {G{\"a}nsicke}, {Hollands}, {Hermes}, {Marsh}, {Cukanovaite}, \&
  {Cunningham}}]{Tremblay19}
{Tremblay}, P.-E., {Fontaine}, G., {Fusillo}, N. P.~G., {et~al.} 2019, \nat,
  565, 202, \dodoi{10.1038/s41586-018-0791-x}

\bibitem[{{van Horn}(1968)}]{VanHorn68}
{van Horn}, H.~M. 1968, \apj, 151, 227, \dodoi{10.1086/149432}

\end{thebibliography}

\end{document}